\documentclass{article}

\usepackage{arxiv}

\usepackage[utf8]{inputenc} 
\usepackage[T1]{fontenc}    
\usepackage{hyperref}       
\usepackage{url}            
\usepackage{booktabs}       
\usepackage{amsfonts}       
\usepackage{nicefrac}       
\usepackage{microtype}      
\usepackage{lipsum}

\usepackage{graphicx}
\graphicspath{ {./} }
\usepackage[
singlelinecheck=false 
]{caption}
\usepackage[para,online,flushleft]{threeparttable}

\usepackage{hyperref}

\makeatletter
\newcommand{\printfnsymbol}[1]{%
  \textsuperscript{\@fnsymbol{#1}}%
}
\makeatother

\title{3D U-Net for segmentation of COVID-19 associated pulmonary infiltrates using transfer learning: State-of-the-art results on affordable hardware}

\author{
  Keno K.~Bressem \\
  Department of Radiology\\
  Charité Universitätsmedizin Berlin\\
  \texttt{keno-kyrill.bressem@charite.de} \\
  
  \And
  Stefan M.~Niehues \\
  Department of Radiology\\
  Charité Universitätsmedizin Berlin\\

  \And
  Bernd ~Hamm \\
  Department of Radiology\\
  Charité Universitätsmedizin Berlin\\

  \And
  Marcus R. ~Makowski \\
  Department of Radiology\\
  Technische Universität München \\

  \And
  Janis L.~Vahldiek\thanks{contributed equally} \\
  Department of Radiology\\
  Charité Universitätsmedizin Berlin\\

  \And
  Lisa C.~Adams\printfnsymbol{1} \\
  Department of Radiology\\
  Charité Universitätsmedizin Berlin

}

\begin{document}
\maketitle

\begin{abstract}
Segmentation of pulmonary infiltrates can help assess severity of COVID-19, but manual segmentation is labor and time-intensive. Using neural networks to segment pulmonary infiltrates would enable automation of this task. However, training a 3D U-Net from computed tomography (CT) data is time- and resource-intensive. In this work, we therefore developed and tested a solution on how transfer learning can be used to train state-of-the-art segmentation models on limited hardware and in shorter time. We use the recently published RSNA International COVID-19 Open Radiology Database (RICORD) to train a fully three-dimensional U-Net architecture using an 18-layer 3D ResNet, pretrained on the Kinetics-400 dataset as encoder. The  generalization of the model was then tested on two openly available datasets of patients with COVID-19, who received chest CTs (Corona Cases and MosMed datasets).  Our model performed comparable to previously published 3D U-Net architectures, achieving a mean Dice score of 0.679 on the tuning dataset, 0.648 on the Coronacases dataset and 0.405 on the MosMed dataset. Notably, these results were achieved with shorter training time on a single GPU with less memory available than the GPUs used in previous studies.
\end{abstract}


\section{Introduction}
The Coronavirus Disease-2019 (COVID-19) is an infectious disease of the respiratory tract and lungs, with more than 80 million confirmed cases worldwide and nearly two million deaths in early 2021 \cite{RN1}. For the management of COVID-19, rapid diagnosis is critical to quickly isolate affected patients and prevent further spread of the disease \cite{RN2}.  \\
Presently, the diagnostic standard for COVID-19 is real-time reverse transcription polymerase chain reaction (RT-PCR) from pharyngeal or deep nasal swaps \cite{RN3}. However, in the clinical setting, computed tomography (CT) is increasingly used in patients with suspected COVID-19. The role of CT to diagnose COVID-19 has been critically debated, and currently there is consensus that CT should not be used in place of RT-PCR \cite{RN4}. Nevertheless, CT remains an important tool for assessing pulmonary infiltrates associated with COVID-19 and for estimating the severity of the disease \cite{RN5}. On CT imaging, COVID-19 typically shows multifocal ground glass opacities as well as consolidations in predominantly peripheral and basal distribution \cite{RN6}. Although the relationship is not strictly linear, a larger affected lung area is associated with more severe disease. Therefore, knowing how much of the lung is affected by COVID-19 may allow a more accurate assessment of disease severity.  \\
Manual segmentation of the affected lung area is a tedious task. In their recent work, Ma et al. manually segmented 20 openly available CT scans of patients affected by COVID-19 an reported a mean duration of 400 minutes per CT volume \cite{RN7}. Clearly, this amount of time is too high to be implemented in routine clinical practice, and research is being conducted on methods to automate these tasks. One of the most promising techniques for automatic segmentation is deep neural networks, in particular the U-Net architecture \cite{RN8}.  \\
U-Nets consist of a down-sampling block that extracts features from input images and an up-sampling part that generates segmentation masks form the previously extracted features. Spatial information decreases in the deeper layers of a convolutional neural network; therefore, the U-Net has skip connections that allow the up-sampling block to use both the feature information of the deeper layers as well as the spatial information from earlier layers to generate high-resolution segmentation masks \cite{RN8}. An advantage of the U-Net architecture is the relatively small amount of data required to obtain accurate results, which is especially important in medical imaging where data are usually sparse \cite{RN8}\cite{RN9}. However, a drawback is the higher memory requirements of the U-Net, since multiple copies of feature maps must be kept in memory to enable the skip connections, so that training a U-Net either requires access to multiple graphics processing units (GPUs) to perform distributed training with a larger batch size, or the batch size must be greatly reduced. This is even more important when U-Nets are extended to three-dimensional space, since each item in a batch of 3D data is even larger. 
Another method to increase the accuracy of a model on limited data is to use transfer learning, where a model architecture is first trained on another task, and then fine-tuned on a novel task \cite{RN10}. \\
 In this work, we developed and evaluated an approach to effectively train a fully three-dimensional U-Net in a single GPU achieving state-of-the-art accuracy by using transfer learning.  

\section{Materials and Methods}
\subsection{Datasets and Annotations}
Three openly available datasets of CT scans from patients affected by COVID-19 are used in this work. These include the following: 
\begin{itemize}
\item RSNA International COVID-19 Open Radiology Database (RICORD) \cite{RN11}
\item MosMedData \cite{RN12}
\item COVID-19 CT Lung and Infection Segmentation Dataset \cite{RN7}
\end{itemize}

RICORD is a multi-institutional and multi-national, expert annotated dataset of chest CT and radiographs. It consists of three different collections:
\begin{itemize}
\item Collection 1a includes 120 CT studies from 110 patients with COVID-19, in which the affected lung areas were segmented pixel by pixel.   
\item Collection 1b contains 120 studies of 117 patients without evidence of COVID-19 
\item Collection 1c contains 1,000 radiographs from 361 patients with COVID-19
\end{itemize}

Only collection 1a was included in the present work. \\
The MosMedData contains data from a single institution. Overall, 1,110 studies are included in the dataset. Pixel-wise segmentation of COVID-19-associated pulmonary infiltrates is available for 50 studies in the MosMedData, which we used for our work. \\
The COVID-19 CT Lung and Infection Segmentation Dataset consists of ten CT volumes from the Coronacases Initiative and ten CT volumes extracted from Radiopaedia, for which the authors have added a pixel-wise segmentation of infiltrates. Because the ten CT volumes extracted from Radiopaedia have already been windowed and converted to PNG (Portable Network Graphics) format, we included only the ten Coronacases Initiative volumes in this study. 

\subsection{Data Preparation}
The RICORD data are provided as DICOM (Digital Imaging and Communications in Medicine) slices for the different CT images, and the annotations are available in JSON format. We used SimpleITK to read the DICOM slices, scale the images according to the rescale intercept and rescale slope, and clip the pixel-values to the range of -2000 and +500 \cite{RN13}. The annotations were converted from JSON (JavaScript Object Notation) to a pixel array and matched to the respective DICOM slice using the study- and SOP instance UID. Both the original volume and annotations were then stored in NIfTI (Neuroimaging Informatics Technology Initiative) format. \\
The MosMedData and COVID-19 CT Lung and Infection Segmentation Dataset were already available in NIfTI format, so no further preprocessing was performed.

\subsection{Model Architecture}
\begin{figure}[h]
  \centering
  \fbox{\rule[-.2cm]{0cm}{0cm}\includegraphics[height=10cm]{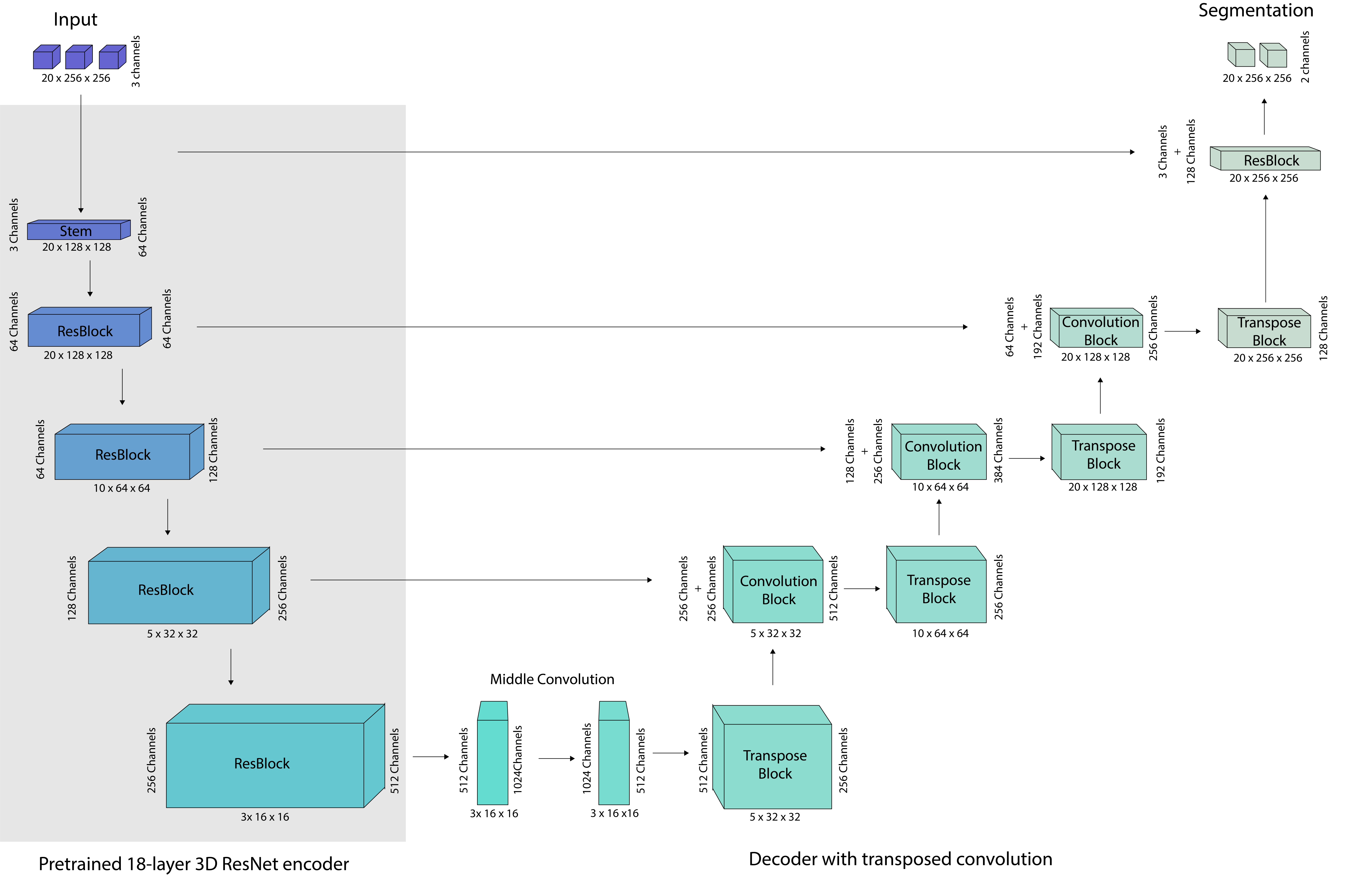}\rule[-.2cm]{0cm}{0cm}}
  \caption{A schematic overview of the network architecture. As the encoder was pre-trained on color images, the expected input size was B x 3 x D x H x W, where B is the batch dimension, D the number of slices and H and W the height and width of each slice. To meet this requirement, the input images were tripled and stacked on the color channel. The encoder consisted out of a basic stem with single convolution, batch normalization and a rectified linear unit. Then, four 3D Residual Block (ResBlock) were sequentially connected to extract the image features. After each ResBlock, a skip connection to the upscaling blocks was implemented. The lower-level features were passed from the last encoder block to a double convolutional layer and then to four sequentially connected upscaling blocks. Each upscaling block consisted of a transposed convolution, which increased the spatial resolution of the feature maps and a double convolutional layer which received the output from the transposed convolution along with the feature maps from the skip connection. The final block of the decoder was again a ResBlock, which reduced the number of feature maps to the specified number of output classes.}
  \label{fig:fig1}
\end{figure}

The 3D U-Net architecture was implemented using PyTorch (version 1.7.0) \cite{RN14} and fastai (version 2.1.10) \cite{RN15}. We used a fully three-dimensional U-Net architecture for CT volume segmentation. The encoder part consisted of an 18-layer 3D ResNet, as described by Tran et al., pretrained on the Kinetics-400 dataset \cite{RN16}. We removed the fully connected layers from the 3D ResNet and added an additional 3D convolutional layer and four upscaling blocks. Each upscaling block consisted of one transposed convolutional layer and two normal convolutional layers. Each convolutional layer was followed by a rectified linear unit (ReLU) as activation function. Instance normalization was applied to the lower layer features before the double convolution was performed. 
The final block of the U-Net consisted of a single residual block without dilation and a single convolutional layer with a kernel size and stride of one for pooling of the feature maps. The model architecture is visualized in the Figure \ref{fig:fig1}.

\subsection{Model Training}

We randomly split the RICORD dataset into a training (85\%) and a tuning (15\%) dataset and used both the MosMedData and COVID-19 CT lung and infection segmentation datasets as hold-out datasets to only evaluate the trained model. \\
A progressive resizing approach was used in which we first trained the U-Net on volumes consisting of 18 slices with a resolution of 112 x 112 px per slice, allowing to use a batch size of 6. In a second training session, we increased the resolution to 256 x 256 px for 20 slices and used a batch-size of 1. \\
During training, we used various augmentations, including perspective distortion, rotation, mirroring, adjusting contrast and brightness, and adding random Gaussian noise to the volumes. For the loss function, we used a combination of the dice loss (as described by Milletari et al. \cite{RN17}) and pixel-wise cross-entropy loss. \\
Regarding the learning rate, we used the cyclic learning rate approach described by Leslie Smith, as implemented in fastai \cite{RN18}. Here, one specifies a base learning rate at the beginning of the training, which is then varied cyclically during each epoch. In addition, the first epochs of the training were warm-up epochs, in which only a fraction of the final learning rate is used. \\
For the first training session, the weights of the pretrained encoder were not allowed to change for the first 10 epochs, and only the randomly initialized weights of the decoder part of the U-Net were trained. To do this, we used a base learning rate of 0.01. We then trained the model for 200 more epochs with a base learning rate of 0.001 and a weight decay of 1e-5. During training, the Dice score on the tuning data was monitored and the checkpoint of the model that achieved the highest dice score was reloaded after training. \\
For the second training session on the higher resolution input data, we set the learning rate to 1e-4 and the weight decay to 1e-5, training for 200 epochs and saving the checkpoint with the highest Dice score. \\
All training was performed on a single GPU (NVIDIA GeForce RTX 2080ti) with 11 GB of available VRAM. 

\section{Results}
The 3D U-Net was trained on the RICORD data (n = 117 CT volumes) which was randomly split into a training dataset consisting out of 100 volumes (85\%) and a tuning dataset of 17 volumes (15\%). The total training duration was 10 hours and 49 minutes with an average duration of 45 seconds per epoch for the lower input resolution and 2:30 minutes for the higher image resolution. While at the beginning of each training session the loss on the training data was higher than on the tuning data, the overall training loss showed a faster decline so that after 200 epochs it was slightly lower than the loss on the tuning data. After 200 epochs, however, we found no obvious signs of overfitting, as the average valid loss was still slowly decreasing

\subsection{Dice score}

The Dice score was used to compare the original segmentation mask with the predicted mask. There are several implementations of the Dice score available that may affect the calculated score and thus limit comparability. We used the implementation by Ma et al., for which the code is freely available \cite{RN7}. \\ 
Because the lung areas affected by COVID-19 can differ substantially from case to case, we calculated the Dice score for each patient and then macro-averaged the scores. This resulted in slightly poorer scores compared with micro-averaging across the entire data set but is more similar to clinical feasibility. \\
We obtained the highest scores on the tuning dataset with a mean Dice score of 0.679 and a standard deviation of 0.13. \\
When applied to new datasets, the performance of the segmentation model decreased with a mean Dice score of 0.648 ± 0.132 for the Coronacases from the COVID-19 CT Lung and Infection Segmentation Dataset, and 0.405 ± 0.213 for the MosMed dataset. A summary of the Dice scores achieved on the datasets is shown in Table~\ref{tab:table1}.

\begin{table}[h]
  \centering
  \caption{Volumetric Dice scores}
  \begin{threeparttable}
  \begin{tabular}{lllll}
    \toprule
    Dataset     & CT scans (n)    & Dice score      & Dice score    & Dice score  \\
                &                 & Mean and std.   & Lowest        & Highest \\
    \midrule
    RICORD	    & 17	          & 0.679 ± 0.130	& 0.398	        & 0.846 \\
    Coronacases	& 10	          & 0.648 ± 0.132	& 0.362	        & 0.783 \\
    MosMedData	& 50	          & 0.405 ± 0.213	& 0.008	        & 0.675 \\
    \bottomrule
  \end{tabular}
  \end{threeparttable}
  \begin{tablenotes}\small
    Overview of the Dice scores obtained for the task of segmenting lung tissue affected by COVID-19 from healthy lung tissue. Abbreviation: Std = standard deviation.
  \end{tablenotes}
  \label{tab:table1}
\end{table}

\subsection{Shape similarity}

Because the normal Dice score is insensitive to shape, we also used the normalized surface Dice (NSD) to assess model performance based on shape similarity \cite{RN19}. To ensure comparability of our results, we again used the implementation of the metric of Ma et al. \cite{RN7}. Again, the highest scores were achieved on the tuning dataset with a mean NSD of 0.781 ± 0.124. On MosMed, the NSD was lowest with a score of 0.597 ± 0.270. On the ten images of the Coronacases dataset, the model achieved an NSD of 0.716 ± 0.135. A summary of the NSD can be found in Table~\ref{tab:table2}. \\
Example images of the segmentation maps generated by the model compared to the ground truth are shown in Figures 2, 3 and 4. Table~\ref{tab:table3}. provides an overview of the results we obtained and those reported in the published literature. 

\begin{table}[htbp]
  \centering
  \caption{Normalized surface Dice scores}
  \begin{threeparttable}
  \begin{tabular}{lllll}
    \toprule
    Dataset     & CT scans (n)    & NSD             & NSD           & NSD  \\
                &                 & Mean and std.   & Lowest        & Highest \\
    \midrule
    RICORD	    & 17	          & 0.781 ± 0.124 	& 0.480	        & 0.911 \\
    Coronacases	& 10	          & 0.716 ± 0.135	& 0.457	        & 0.862 \\
    MosMedData	& 50	          & 0.597 ± 0.270	& 0.060	        & 0.926 \\
    \bottomrule
  \end{tabular}
  \end{threeparttable}
  \begin{tablenotes}\small
    Overview of the achieved normalized surface Dice scores (NSD) as a measurement of shape similarity between two regions. Abbreviation: Std = standard deviation.
  \end{tablenotes}
  \label{tab:table2}
\end{table}

\begin{table}[h]
  \centering
  \caption{Overview of the results from previous studies}
  \begin{threeparttable}
  \begin{tabular}{llllll}
    \toprule
    Publication	                &   Dataset     & Dice score   & Dice score        & Training time     & Hardware \\
                                &               & Tuning data  & Hold-out data     &                   &           \\
    \midrule
    Our approach                & RICORD        & 0.698        &                   & 10h, 49min	       & 1 GeForce RTX 2080ti \\ 
                                & Coronacases   &              & 0.623             &                   & (11GB VRAM) \\
                                & MosMedData	& 	           & 0.403             &                   &               \\
    Müller et al$^1$  \cite{RN9}    & RICORD        & 0.761        & -                 & 130h   	       & 1 Nvidia Quadro P6000 \\ 
                                &               &              &                   &                   & (24 GB VRAM) \\
    Yan et al \cite{RN23}       & proprietary   & -            & 0.726             & -       	       & 6 Nvidia TITAN RTX \\ 
                                &               &              &                   &                   & (24 GB VRAM) \\
    Ma et al$^2$  \cite{RN7}       & Coronacases   & 0.642        & -                 & -       	       &  \\ 
                                & MosMedData    &              & 0.443             &                   &  \\
    Pu et al$^3$  \cite{RN22}        & proprietary   & -            & 0.81          & -       	       &  \\ 
    \bottomrule
  \end{tabular}
  \end{threeparttable}
  \begin{tablenotes}\small
    $^1$ Müller et al. report the accuracy for 5-fold cross-validation; we report the mean of 5 folds. \\
    $^2$ Ma et al. defined different tasks for segmentation, of which we report the accuracy of subtasks 3, as it is the most similar to our methods and thus most comparable. \\
    $^3$ Pu et al. report the Dice score only for lung areas > 200mm$^3$ and rated each infiltration separately. 
  \end{tablenotes}
  \label{tab:table3}
\end{table}

\begin{figure}[h]
  \centering
  \fbox{\rule[-.2cm]{0cm}{0cm}\includegraphics[width=16.2cm]{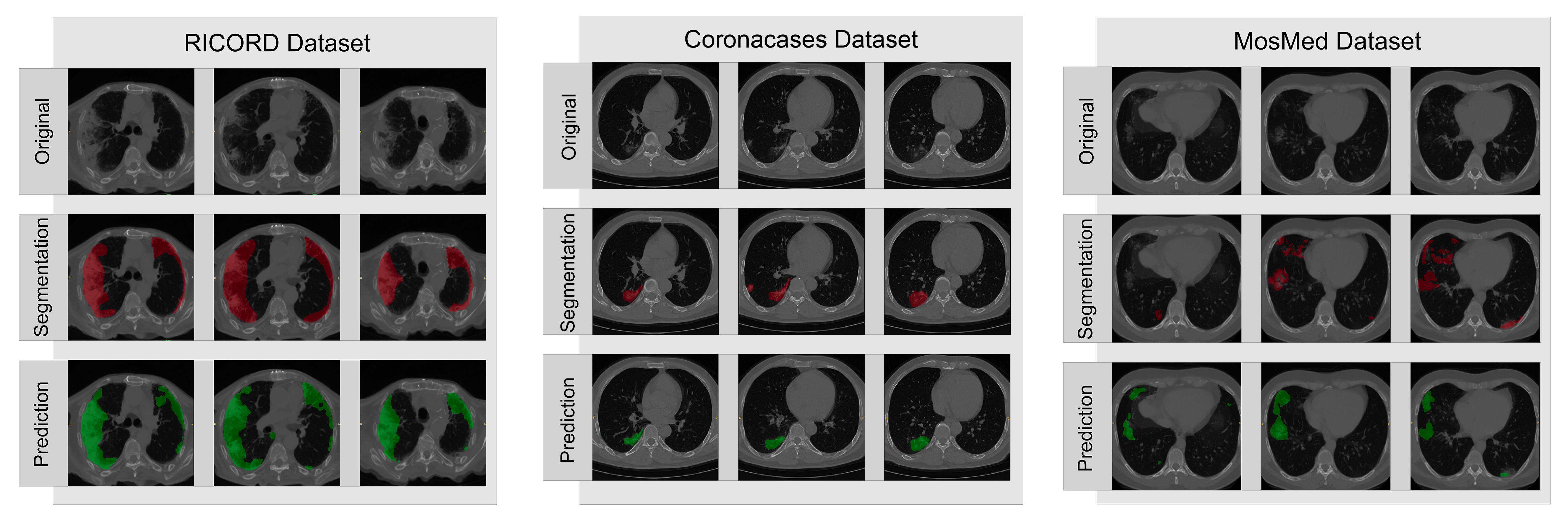}\rule[-.2cm]{0cm}{0cm}}
  \caption{Example images taken from the three datasets used in this study with segmentation masks from a human annotator (red) and the corresponding predicted masks from our model (green). The CT from the MosMed dataset was originally acquired in prone position but images were flipped for this figure. }
  \label{fig:fig2}
\end{figure}

\section{Discussion}
In the present study, we propose a transfer learning approach using a 3D U-Net for segmenting pulmonary infiltrates associated with COVID-19 implemented on a single GPU with 11 GB VRAM. We used a transfer learning approach with an 18-layer 3D ResNet pretrained on a video classification dataset serving as encoder for the 3D U-Net, and obtained state-of-the-art results within comparably short training times. \\
There have been previous efforts to automatically segment pulmonary infiltrates using U-Nets, but few used fully three-dimensional models, while most studies applied a layer-by-layer approach. In our opinion, the metrics obtained from these two approaches are not comparable because the slice-wise approach may introduce selection bias into the data by excluding slices that do not show lung or infiltrates. For 3D models, the input volume shows the entire lung, including healthy and diseased lung tissue, as well as portions of the neck and abdomen that do not contain lung tissue. \\
Müller et al. proposed a fully 3D U-Net, with an architecture similar to our model \cite{RN9}. Because of limited training data, they used 5-fold cross-validation during training and reported a mean Dice score of 0.761 on the 5 validation folds. The model of Müller et al. was trained for 130h (more than 10 times longer than the model presented in this work) on a GPU with twice as much VRAM (Nvidia Quadro P6000). However, since the models were evaluated on a proprietary dataset, the obtained Dice scores cannot be compared without reservations, as differences in segmentation ground-truth may exist. \\
Lessmann et al. developed CORADS-AI, a deep learning algorithm for predicting the CO-RADS grade on non-contrast CT images \cite{RN20}. CO-RADS (COVID-19 Reporting and Data System) is a categorical score between 1-5 that indicates the likelihood of pulmonary involvement, with a CO-RADS score of 1 corresponding to a very low probability of pulmonary involvement and a score of 5 representing a very high probability \cite{RN21}. Interestingly, the interrater agreement on CO-RADS is only moderate, with a Fleiss kappa value of 0.47. CO-RADS grading differs from manual segmentation of pulmonary infiltrates in patients with proven COVID-19 and the kappa values are therefore not transferable. Nevertheless, the question is whether there is also a significant interrater difference in segmentation and how this would affect model performance and comparability between studies. For the RICORD dataset and the dataset provided my Ma et al., each CT volume was annotated by multiple experts, including at least one board-certified radiologist, to reduce bias coming from poor interrater agreement. However, for the MosMed dataset the number of annotators per CT volume is not available. \\
Ma et al. also developed a data-efficient 3D U-Net model that achieved a mean Dice score of 0.642 in the 5-fold cross validation and a Dice score of 0.443 during interference on the MosMed dataset. \\
The highest Dice score achieved with a 3D U-Net architecture was published by Pu et al. with a value of 0.81 for infiltration greater than 200 mm$^3$ on a proprietary dataset \cite{RN22}.  It is important to note, however, that the measurement of Pu et al. differs from other published results as well as from ours because the Dice score is calculated at a per-lesion level and then averaged, rather than at a per-patient level. \\
Yan et al. proposed a novel adaption of the U-Net architecture to increase segmentation performance for COVID-19 \cite{RN23}. Their COVID-SegNet achieved a Dice score of 0.726 on the independent hold-out dataset. To achieve this, they used a proprietary dataset of 861 patients (8 times larger than the RICORD dataset and 40 times larger than the Ma et al.data) and trained their model on six Nvidia Titan RTXs with 24 GB VRAM each. \\
By comparison, the model developed in this study achieved a higher Dice score than Ma et al. and had substantially shorter training times and lower hardware requirements than previously published studies. However, this comparison should be taken with caution because the datasets, training methods and calculation of metrics differed. 
Nonetheless, this study demonstrates the added benefit of using a pre-trained encoder for 3D U-Nets, as one can quickly achieve state-of-the-art results with lower hardware requirements and shorter training times. Transfer learning may help to provide better access and use of 3D segmentation models for the diagnostic community and for researches without access to high performance computing clusters.

\bibliographystyle{unsrt}  
\bibliography{references}  

\end{document}